\documentclass[twocolumn,preprintnumbers,amsmath,amssymb,aps]{revtex4-1}
\usepackage{graphicx}

\begin{document}

\title{Magnetic effects in sulfur-decorated graphene}

\author{Choongyu Hwang,$^{1,2\ast}$ Shane A. Cybart,$^{1,3}$ S. J. Shin,$^{1,4}$ Sooran Kim,$^{5,6}$ Kyoo Kim,$^{5,7}$ T. G. Rappoport,$^{8}$ S. M. Wu,$^{1,3}$ C. Jozwiak,$^{1,9}$ A. V. Fedorov,$^9$ S. -K. Mo,$^9$ D.-H. Lee,$^{1,3}$ B. I. Min,$^{5}$ E. E. Haller,$^{1,4}$ R. C. Dynes,$^{1,3,10}$ A. H. Castro Neto$^{11}$ \& Alessandra Lanzara$^{1,3\ast}$}

\affiliation{$^1$Materials Sciences Division, Lawrence Berkeley National Laboratory, Berkeley, California 94720, USA,}
\affiliation{$^2$Department of Physics, Pusan National University, Busan 46241, Republic of Korea,} 
\affiliation{$^3$Department of Physics, University of California, Berkeley, CA 94720, USA,}
 \affiliation{$^4$Department of Materials Science and Engineering, University of California, Berkeley, CA 94720, USA,}
\affiliation{$^5$Department of Physics, Pohang University of Science and  Technology, Pohang 37673, Republic of Korea,}
\affiliation{$^6$c\_CCMR, Pohang University of Science and Technology, Pohang 37673, Republic of Korea,}
\affiliation{$^7$MPPC\_CPM, Pohang University of Science and Technology, Pohang 37673, Republic of Korea,}
\affiliation{$^8$Instituto de F\'{i}sica, Universidade Federal do Rio de Janeiro, Caixa Postal 68528, 219410972 Rio de Janeiro RJ, Brazil,}
\affiliation{$^9$Advanced Light Source, Lawrence Berkeley National Laboratory, Berkeley, CA 94720, USA,}
\affiliation{$^{10}$Department of Physics, University of California, San Diego, CA 92093, USA,}
\affiliation{$^{11}$Centre for Advanced 2D Materials, National University of Singapore, Singapore 117542,}
\affiliation{$^{12}$Correspondence and requests for materials should be addressed to A.L.~(email: ALanzara@lbl.gov) and C.H.~(email:ckhwang@pusan.ac.kr).}

\begin{abstract}
The interaction between two different materials can present novel phenomena that are quite different from the physical properties observed when each material stands alone. Strong electronic correlations, such as magnetism and superconductivity, can be produced as the result of enhanced Coulomb interactions between electrons. Two-dimensional materials are powerful candidates to search for the novel phenomena because of the easiness of arranging them and modifying their properties accordingly. In this work, we report magnetic effects of graphene, a prototypical non-magnetic two-dimensional semi-metal, in the proximity with sulfur, a diamagnetic insulator. In contrast to the well-defined metallic behaviour of clean graphene, an energy gap develops at the Fermi energy for the graphene/sulfur compound with decreasing temperature. This is accompanied by a steep increase of the resistance, a sign change of the slope in the magneto-resistance between high and low fields, and magnetic hysteresis. A possible origin of the observed electronic and magnetic responses is discussed in terms of the onset of low-temperature magnetic ordering. These results provide intriguing insights on the search for novel quantum phases in graphene-based compounds.
\end{abstract}

\maketitle

Due to the two-dimensional nature and Dirac fermionic behaviour of charge carriers, the electronic and magnetic properties of graphene can be easily modified, making it one of the most appealing materials for a variety of disparate applications~\cite{Geim1,Zhang,Geim2,Oleg}.  Such easiness to access novel regimes has shifted the focus in the graphene research from graphene itself to the modification of graphene, providing an exciting and versatile platform for realization of novel phenomena and device functionality~\cite{Ferrari}. One of such efforts is to induce magnetic effects in graphene. Indeed a growing number of studies ranges from theoretical predictions on the intrinsic ferromagnetism or spin ordering~\cite{Fujita1996,Peres2005,Guinea2005,Son2007,Sarma2007,Ezawa2007,Wang2008} to experimental probes of defect/impurity-induced local magnetic moments~\cite{Wang2009,Xie2011,Esquinazi,Esquinazi2,Ohldag,Barzola,Makarova2000,Narymbetov2000}. However, these magnetic moments are induced by the change of local crystal structure, e.\,g.\,, carbon vacancies and deformations from $sp^2$ to $sp^3$-type crystal structure, and hence, they have been identified as spin-1/2 paramagnets~\cite{Nair2012}. On the other hand, magnetic ordering can be realized in graphene when decorated with sulfur. Upon doping with sulfur, stacked graphene layers, i.\,e.\,, graphite, exhibit ferromagnetism, which has been claimed to coexist with superconductivity~\cite{Silva2001,Moehlecke2004} .

In this work, we have combined two different but complementary probes such as angle-resolved photoemission spectroscopy (ARPES) to study the evolution of the graphene band structure upon sulfur introduction and magneto-transport to explore the electro-magnetic properties of the graphene/sulfur (G/S) system. Figure~1A shows the procedure adopted to prepare the G/S samples. First, graphene samples are grown epitaxially on {\it n}-doped 6{\it H}-SiC(0001) and undoped 4{\it H}-SiC(000\={1}) surfaces by silicon sublimation method, as detailed elsewhere~\cite{Forbeaux,deHeer,XZYu}.  The graphene sample and a piece of sulfur are then sealed in a glass ampule, with a vacuum of $10^{-6}$~Torr. The ampule is annealed at $230~^{\circ}{\rm C}$ in a furnace for 60 hours, while the pressure inside the ampule increased by vapourized sulfur was $\sim$360~Torr. The presence of sulfur in the samples is confirmed by the observation of: a) sulfur 2{\it p} core electrons in the photoemission spectra and b) Auger electrons corresponding to the sulfur {\it LMM} transition in the Auger electron spectroscopy (AES) spectra~\cite{Gallon}. The concentration of sulfur is determined by AES and the value of sulfur/carbon ratio is 1/9.

\begin{figure}
  \begin{center}
  \includegraphics[width=1\columnwidth]{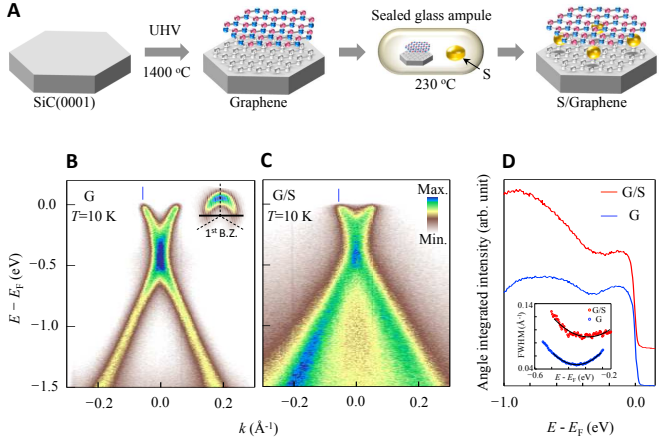}
  \end{center}
\caption{Sample preparation process and energy spectra near $E_{\rm F}$. (A) An SiC substrate is annealed at 1400~$^{\circ}$C in ultra-high vacuum (UHV) to grow graphene, followed by another annealing process with a piece of sulfur in a sealed glass ampule at 230~$^{\circ}$C. (B-C) ARPES intensity maps of the as-grown sample (G: panel (B)) and the graphene/sulfur compound (G/S: panel (C)), both of them measured at 10~K. The inset in panel (B) is the Fermi surface of the G sample where the black line is the direction that the energy-momentum dispersion of both samples has been taken. (D) Angle-integrated intensity of the energy spectra for the G (blue curve) and G/S (red curve) samples shown in panels (B) and (C), respectively. The inset shows full width at half maximum (FWHM) of the momentum distribution curves (MDCs) near the Dirac energy for G (blue circles) and G/S (red circles) samples. Both spectra show that the minimum corresponding to $E_{\rm D}$ shifts towards $E_{\rm F}$ upon sulfur introduction.} \label{Fig1}
\end{figure}

Changes in the electronic structure are monitored by ARPES experiments. Figures~1B and~1C show ARPES intensity maps as a function of energy and momentum for as-grown graphene (G) and G/S samples. Data are taken at 10~K, near the Brillouin zone corner K, along the black line shown in the inset. In line with previous reports~\cite{Rollings,OhtaPRL,Zhou2007NM}, the Dirac energy, $E_{\rm D}$, of the G sample lies $\sim$0.35~eV below the Fermi energy, $E_{\rm F}$, due to the formation of a Schottky barrier~\cite{Seyller}. Two most obvious effects resulting from the introduction of sulfur are: a) broadening of the ARPES spectra; and b) small shift of $E_{\rm D}$ towards $E_{\rm F}$. The former suggests that sulfur introduces disorder and the latter indicates charge transfer between sulfur and the graphene layer. More specifically, $E_{\rm D}$ is estimated by the minimum of the integrated ARPES intensity (Fig.~1D), a quantity proportional to the one-dimensional density of states, or the full width at half maximum (FWHM) of the momentum distribution curves (MDCs, momentum spectra at constant energy shown in the inset of Fig.~1D) that shifts towards $E_{\rm F}$ by $67\pm6~{\rm meV}$ with introduction of sulfur. This signifies a decrease of the carrier concentration by 17~\% from $0.90\times10^{13}~{\rm cm}^{-2}$ to $0.75\times10^{13}~{\rm cm}^{-2}$.

\begin{figure}
  \begin{center}
  \includegraphics[width=1\columnwidth]{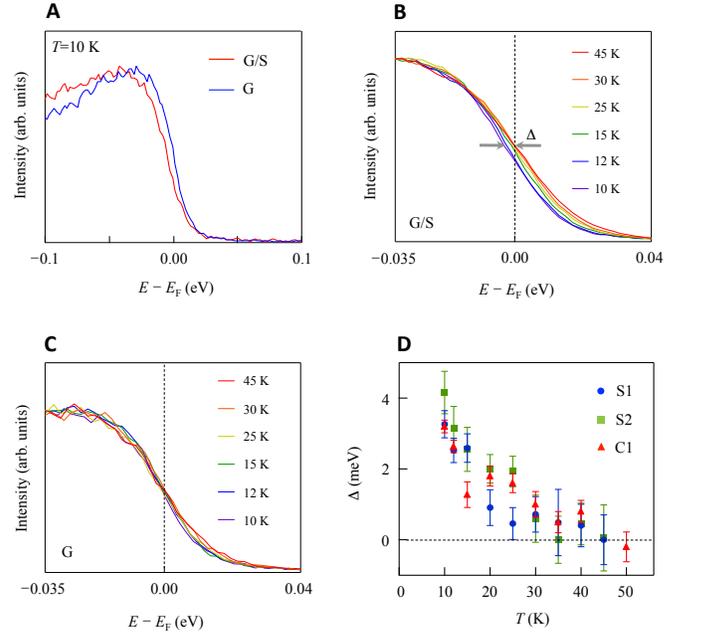}
  \end{center}
\caption{Energy gap at $E_{\rm F}$ of G/S. (A) Energy distribution curves (EDCs) taken at 10~K for G (blue curve) and G/S (red curve) samples. (B) EDCs taken at several temperatures for G/S. The leading edge shifts away from $E_{\rm F}$ with decreasing temperature. The leading edge gap $\Delta$ (roughly half the energy gap) is determined by the position of the leading edge with respect to $E_{\rm F}$. (C) EDCs taken at several temperatures for G. The leading edge stays at the same energy within the fitting error ($\sim$$\pm$0.6~meV). (D) Temperature dependence of $\Delta$ for three different samples: S1 and S2 (G/S on an SiC(0001) substrate), and C1 (G/S on an SiC(000$\bar{1}$) substrate).} \label{Fig1}
\end{figure}

A closer look at the energy distribution curves (EDCs, energy spectra at constant momentum $k_{\rm F}$ indicated by the vertical blue lines in Figs.~1B and~1C) reveals a surprising depletion of states at $E_{\rm F}$ (Fig.~2A). This is a typical signature of energy gap opening at $E_{\rm F}$. The temperature dependence of the EDCs for the G/S sample is shown in Fig.~2B. As the temperature $T$ is lowered, a clear depletion of states at $E_{\rm F}$ or a shift of the leading edge toward higher binding energy is observed. This is clearly in striking contrast to the G sample (Fig.~2C), where, as previously reported~\cite{Seba}, no gap is observed when measured below 45~K. The magnitude of the energy gap is determined by the leading edge of the EDCs with respect to $E_{\rm F}$ through the standard procedure~\cite{Rickert} of fitting the first derivative of each energy spectrum with a Gaussian function. The temperature dependence of the gap is summarized in Fig.~2D. The gap develops below a transition temperature $T_{\rm c}\sim30\pm10~{\rm K}$ with a maximum shift of $4.1\pm0.6~{\rm meV}$ at 10~K and appears to be independent from the substrate. The different $T$-dependence of G/S from G excludes the possibility of thermal smearing as the origin of the depletion of states at $E_{\rm F}$. Several mechanism can be accounted for the gap opening from superstructures~\cite{Pletikosic2009,Park1,Park2} and charge- or spin-density wave~\cite{Gruner} to magnetic ordering~\cite{Rappoport2011} and superconductivity~\cite{Rose}. The lack of replica (shadow) bands corresponding to the additional periodicities, however, makes the formation of the former two less likely scenarios. Indeed, considering the sulfur/carbon ratio present in our G/S sample, one would expect the gap opening at a momentum of 0.3$\sim$0.4~\AA$^{-1}$, which is far away from 0.05~\AA$^{-1}$ where the energy gap is observed (Fig.~1C). Magnetism and superconductivity are certainly appealing scenarios also in view of the report of coexisting ferromagnetism and superconductivity in sulfur-doped graphite~\cite{Silva2001}.

\begin{figure}
  \begin{center}
  \includegraphics[width=1\columnwidth]{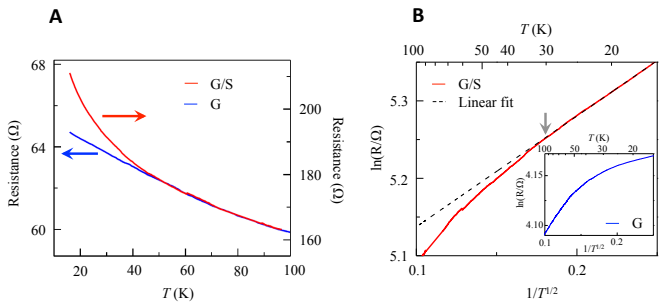}
  \end{center}
\caption{Transport properties of G/S. (A) The $R$ versus $T$ curves of G (blue curve) and G/S (red curve) samples. (B)  The $\ln$($R$) versus $T^{-1/2}$ curve of the G/S sample. The black-dashed line is a $R=C{\rm \exp}(T_0/T)^{1/2}$ fit, where $T_0=$~2.49~K and $C$ is an arbitrary constant. The inset shows the $\ln$($R$) versus $T^{-1/2}$ curve of the G sample, for comparison.} \label{Fig1}
\end{figure}

To gain better insight on the origin of the gap opening, we have performed magneto-transport measurements (Figs.~3 and~4). In each case, the introduction of sulfur strongly modifies the response of the G sample. Especially, the resistance $R$ of G/S gradually deviates and sharply increases with respect to the one of the G sample with lowering $T$ (Fig.~3A). Figure~3B displays a $\ln$($R$) versus $T^{-1/2}$ curve for the same data~\cite{Yu}. The linear $T$-dependence observed at low temperatures is typical of disordered systems in the variable range hopping regime (VRH)~\cite{Yu,Joung,Chuang}, consistent with an increase in the disorder of the G/S sample as discussed in Fig.~1C. This is qualitatively different from that of G (inset of Fig.~3B), where the linearity is not well defined due to the curvature in the whole $T$ range.

\begin{figure}
  \begin{center}
  \includegraphics[width=1\columnwidth]{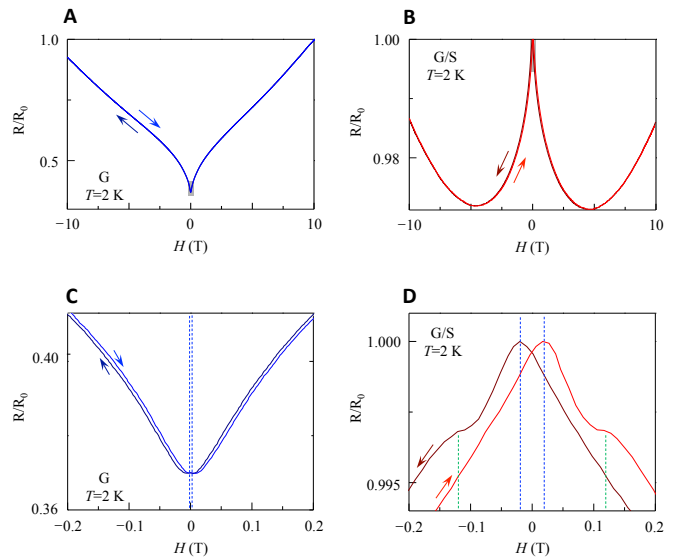}
  \end{center}
\caption{Magnetic hysteresis of G/S. (A) Magneto-resistance at 2~K of the G sample, with increasing (blue curve) and decreasing (dark blue curve) magnetic field, $H$. R$_0$ is the maximum resistance (91~$\Omega$) at $H$=10~T. (B) Magneto-resistance at 2~K for the G/S sample, with increasing (red curve) and decreasing (dark red curve) $H$. R$_0$ is the maximum resistance (6.05~k$\Omega$) near $H$=0~T. (C) The magnetoresistance of G near $H$=0~T denoted by gray-shaded area in panel~(A). The hysteretic effect with a coercive field of 0.003~T (blue-dashed lines) is observed. (D) The magnetoresistance of G/S near $H$=0~T denoted by gray-shaded area in panel~(B), showing magnetic hysteresis with two coercive fields of 0.12~T (green-dashed lines) and 0.02~T (blue-dashed lines).} \label{Fig1}
\end{figure}

In Fig.~4, we show the low-temperature magneto-resistance ($R$ versus external magnetic field $H$) for both samples. The most striking difference is the behaviour of the magneto-resistance at low fields, where its complete reversal is observed upon sulfur introduction. This reversal can be explained with a transition from a weak antilocalization regime to a weak localization regime as previously reported for pure graphene~\cite{Wu,Tikhonenko}. For this case, intervalley scattering is turned on, thus causing an increase in $R$ near zero field. While the magneto-resistance of the G sample (Fig.~4A) is positive (increasing $R$ upon increasing $H$) over the whole field range, the magneto-resistance of the G/S sample (Fig.~4B) shows a crossover at $\sim$5~T from positive at high fields to negative (decreasing $R$ upon increasing $H$) at low fields, similar to the case of weak localizations. We note, however, that the crossover field is three orders of magnitude higher than the previous results for weak localizations, where the strength of localization is suppressed very quickly in the presence of magnetic field~\cite{Wu}. Instead, the negative magneto-resistance at low fields is reminiscent of fluorinated graphene~\cite{Hong2011}, identified as a spin-1/2 paramagnet~\cite{Nair2012}.

The magneto-resistance near $H$=0~T reveals another surprising phenomena of G/S, magnetic hysteresis with coercive fields of 0.12~T and 0.02~T (defined to be half the distance between the two peaks denoted by green and blue dashed lines in Fig.~4D). On the other hand, the G sample measured under the same condition (Fig.~4C) shows one order of magnitude smaller hysteretic effect with a coercive field of 0.003~T, likely induced by a remnant field of the superconducting magnet used for the measurements. Hysteresis in the magneto-resistance normally indicates the formation of magnetic domains or the existence of a magnetic granular system. The domains or grains change polarity above a given coercive field, creating a remanence in the signal. In the case of the G/S sample, the observed hysteresis persists in high fields up to 5~T, suggesting very large magnetic anisotropy and/or saturation fields. The observed magnetic hysteresis in the disordered system (broad ARPES spectrum in Fig.~1C along with the $R\sim{\rm exp}^{1/T^{1/2}}$ behaviour in Fig.~3B) requires that magnetic moments are correlated with electronic states of graphene in the presence of sulfur. 

\begin{figure}
  \begin{center}
  \includegraphics[width=1\columnwidth]{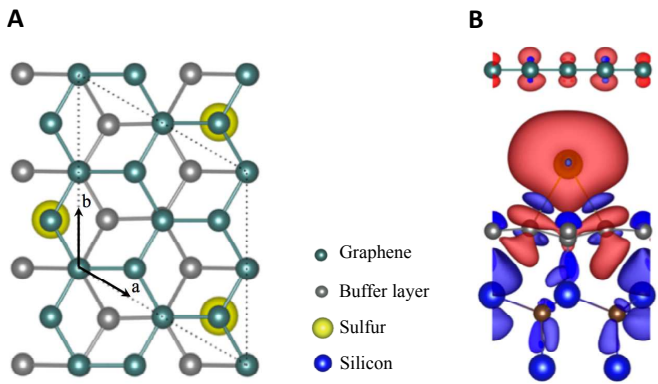}
  \end{center}
\caption{Magnetization of an intercalated sulfur atom. (A) The top view of the crystal structure of G/S (the SiC substrate is not shown for simplicity). The dotted lines denote the unit cell of G/S compared to the unit cell of graphene. (B) The spin density of G/S. The red and blue iso-surfaces are for spin up and down, respectively.} \label{Fig1}
\end{figure}

One of the plausible origins of the observed magnetic moments is magnetized sulfur atoms. Our first principles calculations for G/S (Fig.~5A) suggest that a non-magnetic sulfur atom, after being trapped in between graphene and the buffer layer (a carbidic layer in between graphene and the SiC substrate whose crystal structure is the same as graphene, but $\pi$ bands are absent due to the interactions with the substrate) exhibits spin-polarized states via the interaction with the buffer layer (see the spin density map in Fig.~5B, where the red and blue iso-surfaces correspond to spin up and down, respectively) exhibiting a magnetic moment of 0.63~$\mu$B. This can result in indirect exchange interactions analogous to the Ruderman-Kittel-Kasuya-Yosida interactions in metals driving the spin-dependent VRH~\cite{Foygel2003,Rappoport2009}. Within this picture, spin-dependent hopping between neighbouring sublattices costs an energy $J$ leading to the opening of a gap in the electronic spectra below a certain temperature~\cite{Rappoport2009,Rappoport2011} similar to the double-exchange mechanism in manganites, i.\,e.\,, magnetism induced by adatoms with a concentration of $n_a$ can open a charge gap $\Delta=2n_a J$ that is equal to $2\times{\rm (leading\,edge\,gap)}$, where $J$ is the strength of exchange interactions~\cite{Daghofer2010}. Indeed, we obtain  $T_{\rm c}=(4.8\pm1.6)J^2/t$ (with $T_{\rm c}\sim30\,{\rm K}=2.6\,{\rm meV}$ and the nearest neighbour hoping parameter $t=3.1\,{\rm eV}$~\cite{Gruneis2008}), when $\Delta=0.2J=8.2~{\rm meV}$ (at 10~K from the results in Fig.~2 with $n_a\approx10~\%$ consistent with C:S=9:1 that we have determined by Auger electron spectra). This result is similar to $T_{\rm c}\approx J^2/t$ what the double-exchange mechanism expects~\cite{Dagotto2001}.  Such spin-dependent VRH and energy gap opening are exactly demonstrated in the temperature-dependence of magneto transport signals (Figs.~3 and~4) and energy spectra (Figs.~2B and~2D).

It is interesting to note that similar behaviour has been observed at the interface between LaAlO$_3$ and SrTiO$_3$~\cite{Brinkman2007,Ariando2011}. This interface is a quasi-two-dimensional metallic layer that presents ferromagnetic order and large negative magneto-resistance~\cite{Brinkman2007}. In this case, electronic charge separation has been suggested as the main ingredient for the appearance of the novel behaviour~\cite{Ariando2011}. At the G/S sample, it is clear that sulfur removes electrons from the graphene layer (Fig.~1D). This similarity suggests that depending on the nature of the charge transfer between sulfur and carbon, nanoscale inhomogeneities are created, leading to nanoscopic droplets with different electronic and magnetic properties, similar to the LaAlO$_3$/SrTiO$_3$ interfaces~\cite{Ariando2011,Bert2011,Kalisky2012}. However, we should stress the stark contrast between these layered oxides and G/S: G/S is metallic not insulating, and hence electronic screening should be more efficient~\cite{review}, Coulomb interactions weaker, and therefore the observed behaviour is even more surprising. 

In conclusion, we have revealed an unusual response to temperature and magnetic field from a material that consists of non-magnetic and light elements such as carbon and sulfur. The results here presented place graphene/sulfur as another appealing system where magnetic ordering can be realized in graphene and carbon-based materials in general, and suggest the appealing possibility to realize novel phenomena such as ferromagnetic quantum Hall effect~\cite{Brey1995} for graphene-based spintronic devices operated not only by temperature and doping, but also by magnetic field.
 
\vspace{\baselineskip}
 
{\bf Methods}\\
Experiments: High-resolution ARPES experiments were performed at beamlines 10.0.1 and 12.0.1 of the Advanced Light Source at Lawrence Berkeley National Laboratory using 50~eV photons, with energy and momentum resolutions of 9~meV and 0.01~\AA$^{-1}$, respectively. The magneto-transport properties were measured in a Van-der-Pauw geometry for the G/S and G samples using a custom-built cryogenic vacuum probe equipped with a 10~T superconducting magnet.\\ \\
Calculations: \textit{Ab initio} total energy calculations were performed with a plane-wave basis set~\cite{Cohen} using the Vienna Ab-initio Simulation Package (VASP) \cite{PhysRevB.47.558,Kresse1996, PhysRevB.54.11169} . The exchange-correlation of electrons was treated within the generalized gradient approximation (GGA) as implemented by Perdew-Berke-Enzelhof~\cite{Perdew}. The  crystal structure is simulated by using the supercell slab approach with vacuum region which separates the two surfaces (top and bottom surfaces). The bottom one is passivated by hydrogens to remove the lone pair of the SiC substrate. The {\it k} point sampling is (9$\times$9$\times$1). The structural optimisation and spin-polarized calculation were performed.  

\acknowledgments  CH, CJ, AL acknowledge financial support from the National Science Foundation grant number DMR-1410660. CH acknowledges also partial financial support from the National  Research Foundation of Korea (NRF) grant funded by the Korea government (MSIP) (No.~2015R1C1A1A01053065), the Research Fund Program of Research Institute for Basic Sciences, Pusan National University, Korea, 2013, Project No. RIBS-PNU-2013-311, and Max Planck Korea/POSTECH Research Initiative of the National Research Foundation (NRF) funded by the Ministry of Science, ICT and Future Planning under Project No.~NRF-2011-0031558. AHCN acknowledges the National Research Foundation, Prime Minister Office, Singapore, under its Medium Sized Centre Programme and CRP award ?Novel 2D materials with tailored properties: Beyond graphene? (R-144-000-295-281). Work at the ALS is supported by DOE BES under Contract No. DE-AC02-05CH11231.

\end{document}